\documentclass[12pt]{article}
\usepackage{amsfonts}
\usepackage[dvips]{graphicx}
\begin{document}
\thispagestyle{empty}
 \vspace*{3cm}
\begin{center}
 \textbf{\Large Heisenberg
uncertainty principle
\\for thermal response of the microtubules \\excited by
ultra-short
laser pulses}

\bigskip
\bigskip
{\Large Janina Marciak-Kozlowska, Miroslaw Kozlowski\footnote{Corresponding author}}\\

Institute of Electron Technology\\

Al. Lotnik\'ow 36/38, Warsaw, Poland

\bigskip
\bigskip
{\Large Magdalena Pelc}

Warsaw University, Science Teachers College

Smyczkowa 5/7, Warsaw, Poland
\end{center}

\vspace{3cm}
\begin{abstract}
In this paper the heat signaling in microtubules (MT) is
investigated. It is argued that for the description of the heat
signaling phenomena in MT, the hyperbolic heat transport (HHT)
equation must be used. It is shown that HHT is the Klein-Gordon
(K-G) equation. The general solution for the K-G equation for MT
is obtained. For the undistorted signal propagation in MT the
Heisenberg uncertainty principle is formulated and discussed.

\end{abstract}
\newpage
\section{Introduction}
The aim of this paper is to provide a semi-quantum theory of
intracellular heat transport of cell organelles and vesicles here
termed ``particles''. Numerous experimental studies have
established that this particles are equipped with bound motor
proteins which move them along microtubules and actin
filaments~\cite{1}. For example anterograde transport of particles
along microtubules in nerve axons is mediated by the motor protein
kinesin~\cite{2}. In this system the motion of particles is not
continuous but saltatory~\cite{3}: particles are transported for
distances of typically $\sim 10$~nm at a more or less steady
velocity of $\sim1\mu{\rm ms}^{-1}$ but there are pauses lasting
for upward of 1~s in which given particle is apparently undergoing
Brownian motion and has presumably detached from the microtubule
or is stuck.

Recently D.~A.~Smith and R.~M.~Simmons~\cite{4} developed the
reaction -- diffusion model for the motor assisted transport of
intracellular particles. In this paper we propose the model for
signaling phenomena inside the microtubules. Considering our
results collected in the monograph~\cite{5} we argue that the
transport phenomena \textit{inside} the microtubules must be
discussed in the frame of the \textit{quantum} transport equations
for the diameter of the microtubules is of the order of
nanometers, i.e. the order of the molecules. In the paper we
develop the Klein-Gordon (K-G) type transport equation and will
obtain the solution  of K-G for Cauchy boundary conditions.

Recently~\cite{10} the study of the motion of the kinesin motor
along microtubules using \textit{interference total internal
reflection microscopy} was undertaken. This technique bears some
similarity to earlier ones designed to study the proximal
hydrodynamics behavior of polymers to a surface. The laser source
wavelength currently used in the experiment is 532 nm. The two
laser beams identical in divergence, intensity phase and
polarization are obtained using a flat beam splitter. From
experiment~\cite{10} the value of the diffusion coefficient of the
bead along the microtubule is obtained, $D=315 {\rm nm}^2{\rm
s}^{-1}$.

In this paper we study the heat signaling phenomena inside the
microtubule excited by ultra-short laser pulses.
\section{Heat pulse transport on the molecular scale}
Molecular electronics is a~new field of science and technology,
which is evolving from the  convergence of ideas from chemistry,
physics, biology, electronics and information technology.
 It considers, on the one
hand molecular materials for electronic/optoelectronic applications, on the
other hand attempts to build electronics with molecules at the
molecular level. It is this second viewpoint which concerns us here: describing
 the heat transport at the level of single or few molecules.

The heat and charge transport phenomena on the molecular scale are
the quantum phenomena and electrons constitute the charge and heat
carriers. We argue, that to describe the heat transport phenomena
 on the molecular level the quantum
heat transport equation is a~natural reference equation. The
quantum heat transport equation  for the atomic scale
 has the form
\begin{equation}
\tau\frac{\partial^2 T}{\partial t^2}+\frac{\partial T}{\partial t}=
D\nabla^2T,\label{eq14:1}
\end{equation}
where $\tau$ is the relaxation time, $D$ is the heat diffusion
coefficient and $T$ denotes temperature. For the atomic scale heat
transport, the relaxation time equals
\begin{equation}
\tau=\frac{\hbar}{mv^2_h},\qquad v_h=\frac{1}{\sqrt{3}}\alpha c,\label{eq14:2}
\end{equation}
where $m$ is the electron mass and $v_h$ is the velocity of the heat
perturbation.  Moreover, on the
atomic level the temperature field $T(r)$ is quantized by a~quantum heat of
energy, the {\it heaton}. The {\it heaton} energy equals
\begin{equation}
E_h=m_ev^2_h.\label{eq14:3}
\end{equation}
Due to formula~(\ref{eq14:3}), the {\it heaton} energy is
 the interaction energy
of electromagnetic field with electrons (through the coupling constant
$\alpha$).

At the molecular level we seek  energy of interactions of the
electromagnetic field with a~molecule. This energy is described by
the formula
\begin{equation}
E^m_h=\alpha^2\frac{m_e}{m_p}m_ec^2,\label{eq14:4}
\end{equation}
where $m_e, m_p$ are the masses of electron and proton, respectively.

Considering the general formula for {\it heaton} energy~(\ref{eq14:3}), one obtains from
formula~(\ref{eq14:4}) for velocity of the thermal
perturbation
\begin{equation}
v_h=\alpha c\left(\frac{m_e}{m_p}\right)^{1/2}.\label{eq14:5}
\end{equation}
Comparison of formulas~(\ref{eq14:2}) and (\ref{eq14:5}) shows, that $v_h$
scales with
ratio~$(m_e/m_p)^{1/2}$ when the atomic scale is changed to
the molecular scale; $v_h$ is the Fermi velocity\index{Fermi velocity!for
molecular gas} for molecular gas.

Quantum heat transport equation~(\ref{eq14:1}) has as a~solution, for short time
scale, (short in comparison to relaxation time $\tau$)
 the heat waves which
propagate with velocity $v_h$. One can say that on the molecular level, the heat
waves are slower in comparison to the atomic scale.

From formulas~(\ref{eq14:2}) and~(\ref{eq14:5}), the relaxation
time  can be calculated
\begin{equation}
\tau=\frac{m_p}{m_e}\frac{\hbar}{m_ec^2\alpha^2}.\label{eq14:6}
\end{equation}
It occurs, that  relaxation time on the molecular
scale is longer (ratio~$m_p/m_e$) than the atomic relaxation time. For
standard values of the  constants of the Nature
 \begin{eqnarray}
\alpha=\frac{1}{137}, \qquad m_e=0.511 \,{\rm MeV/c^2}, \qquad m_p=938\,
 {\rm MeV/c^2},\label{eq14:7}
\end{eqnarray}
one obtains the following numerical values for $v_h$, $\tau$ and $E_h$: $
v_h=0.05\,
{\rm nm/fs}, \newline \tau=44 \,{\rm fs}$ and $E_h=10^{-2}\;{\rm eV}$. With those
values of $v_h$ and $\tau$, the
mean free path
\begin{equation}
\lambda=\tau v_h\label{eq14:8}
\end{equation}
can be calculated and $\lambda=2.26 \,{\rm nm}$. It is interesting
to observe, that in the structure of the biological cells (e.g.
microtubules), some elements have the dimension of the order of
the nanometer (e.g. microtubules).

With the help of the {\it heaton} energy one can define the {\it heaton
 temperature},\index{Heaton temperature}~i.e.,~the
characteristic temperature of the heat transport on the molecular scale, viz.:
\begin{equation}
T_m=\alpha^2\frac{m_e}{m_p}m_ec^2\,1.16\, 10^{10}\, {\rm K}
\approx10^{-3}\alpha^2m_ec^2\sim316\, {\rm K}.\label{eq14:9}
\end{equation}
This defines what we generally term ``room temperature''. At
temperatures far below $T_m$, the hydrogen bond becomes very rigid
and the flexibility of atomic configurations is weakened. Most
substances are liquid or solid below $T_m$. Biology occurs in
environments with ambient temperature within an order or magnitude
or so of $T_m$. {\linespread{1}
\begin{figure}[p]
\begin{center}
\begin{picture}(227,214)
\put(0,-40){\includegraphics{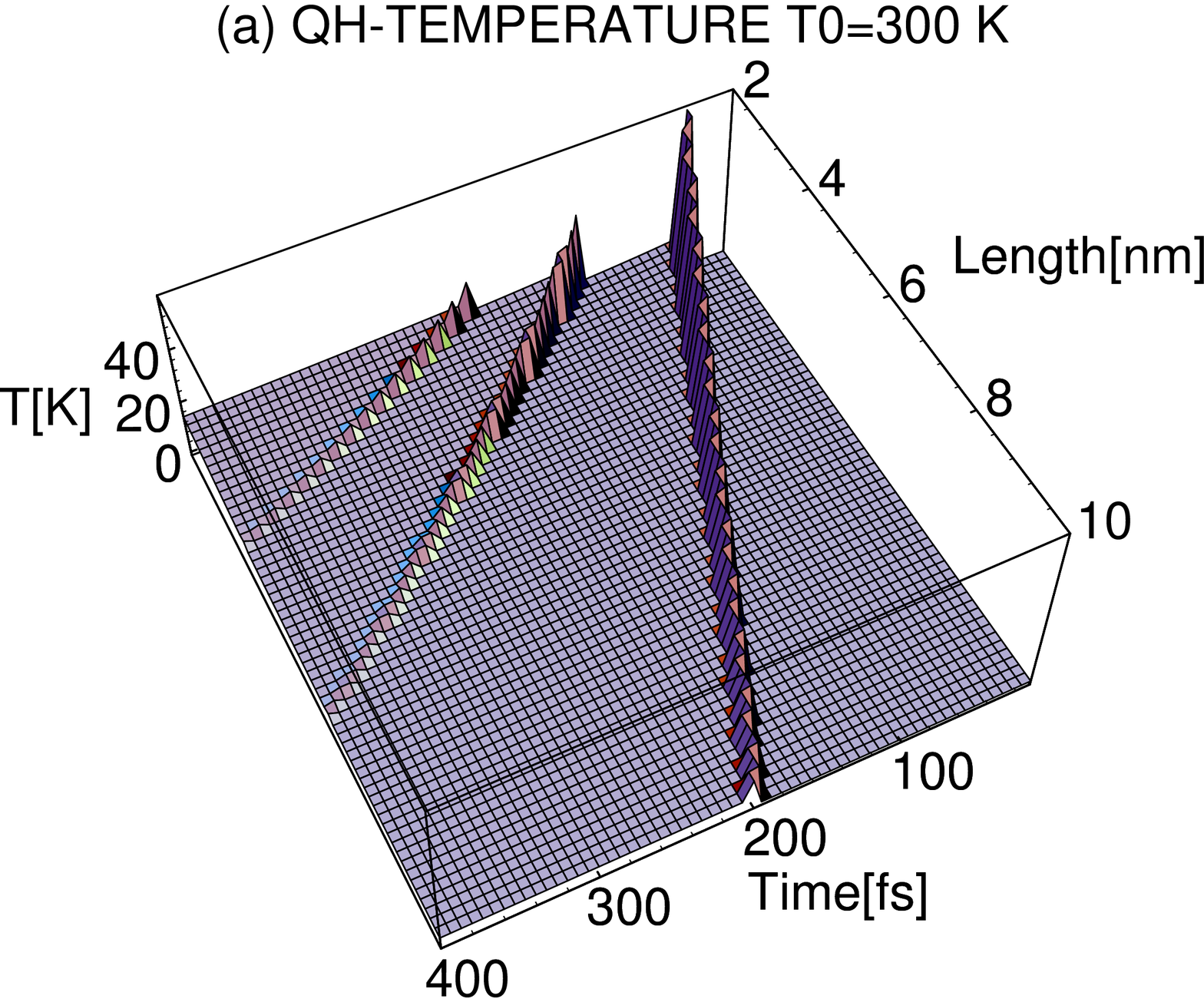}}
\end{picture}
\begin{picture}(227,214)
\put(0,-40){\includegraphics{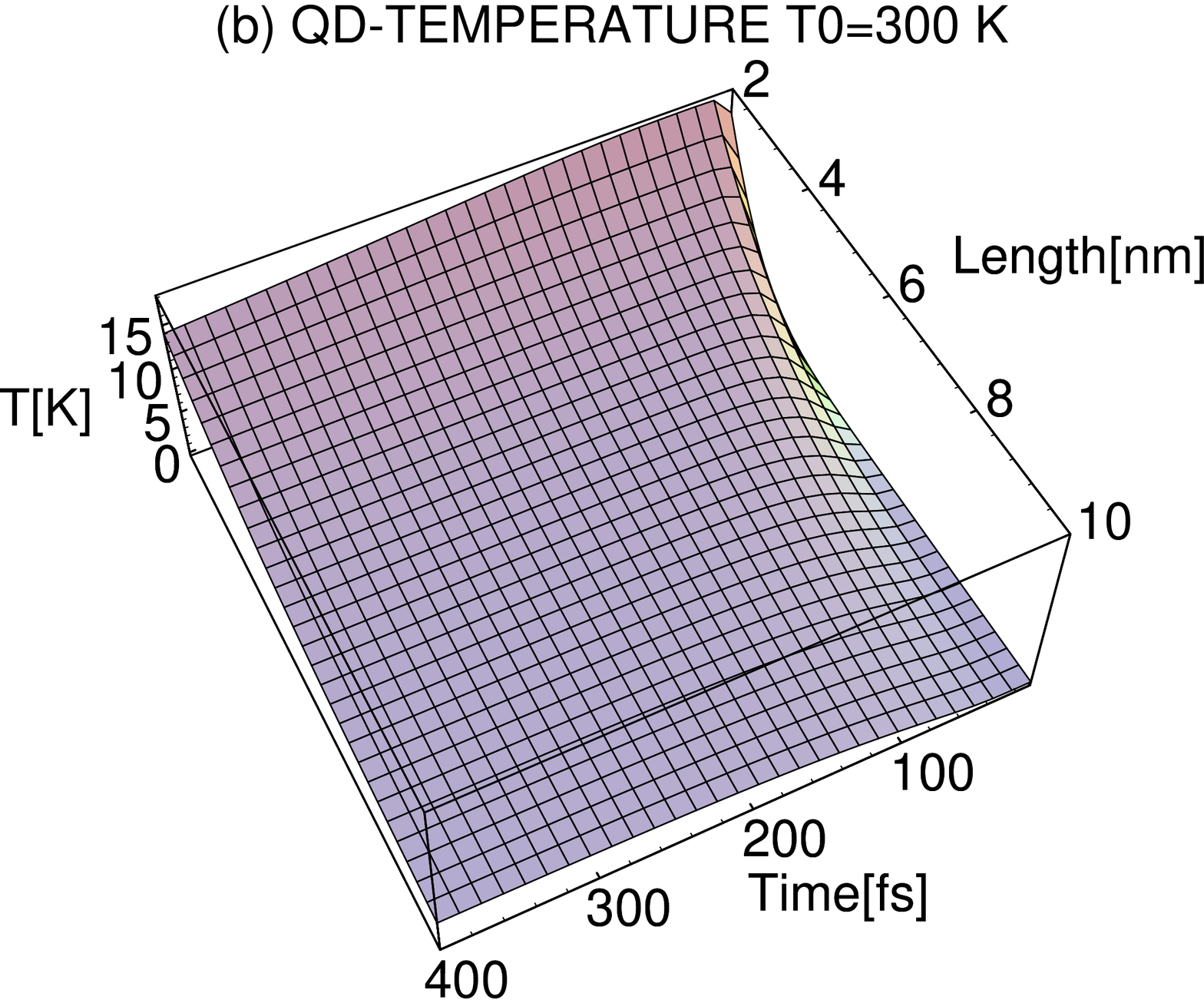}}
\end{picture}
\end{center}
\caption{(a) The solution of QHT equation~(\ref{eq14:1}) for the
following input parameters $v_h=5\; 10^{-2} \,{\rm nm/fs}, \tau=44
\, {\rm fs}, T_0=300 \,
 {\rm K}$  and
$\Delta t=0.2\tau$. (b)~The~solution of QPT~(\ref{eq14:11}) with
the same input parameters.\label{fig14.1}}
\end{figure}

\begin{figure}[p]
\begin{center}
\begin{picture}(241,228)
\put(20,-20){\includegraphics{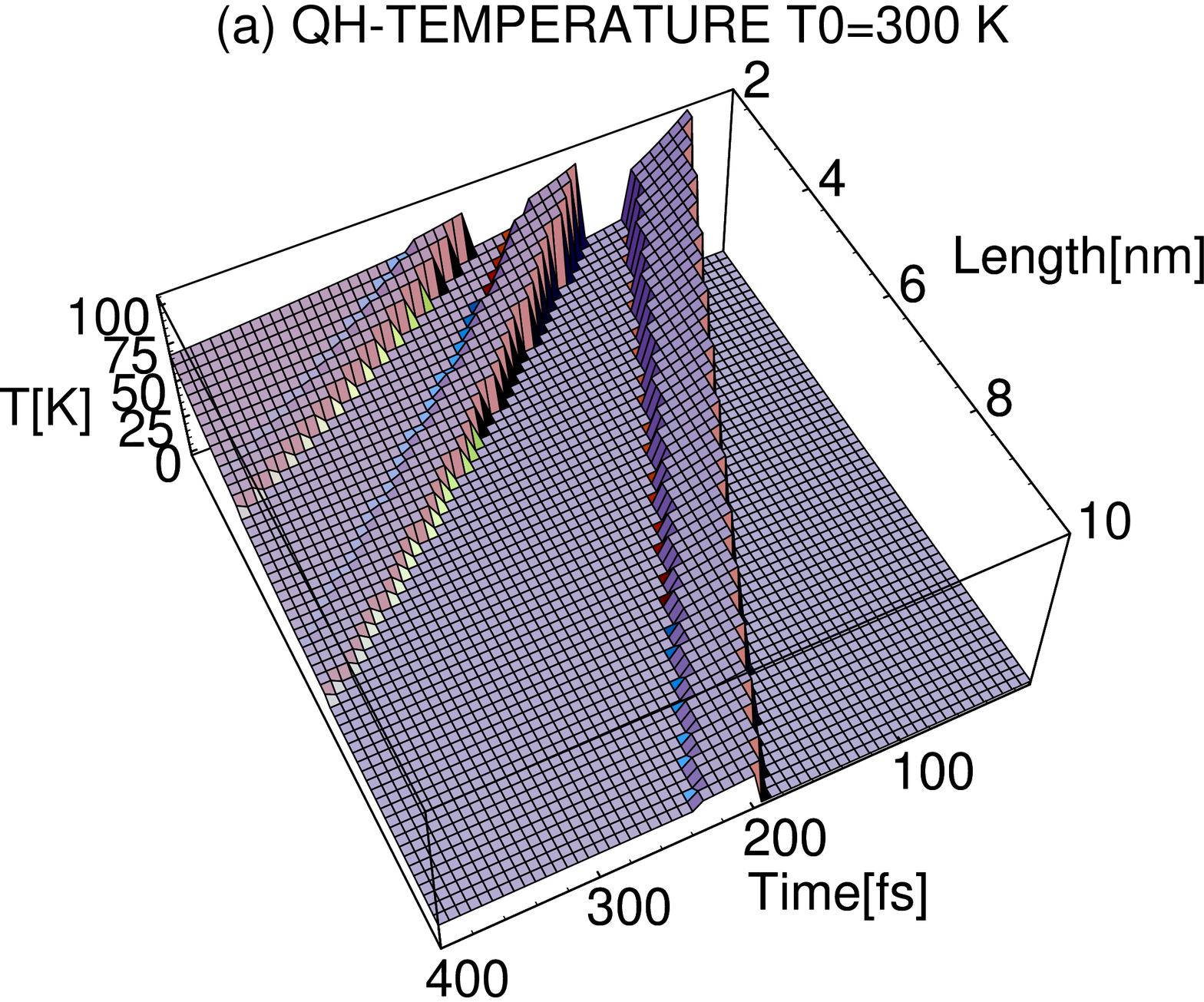}}
\end{picture}
\begin{picture}(241,228)
\put(20,-20){\includegraphics{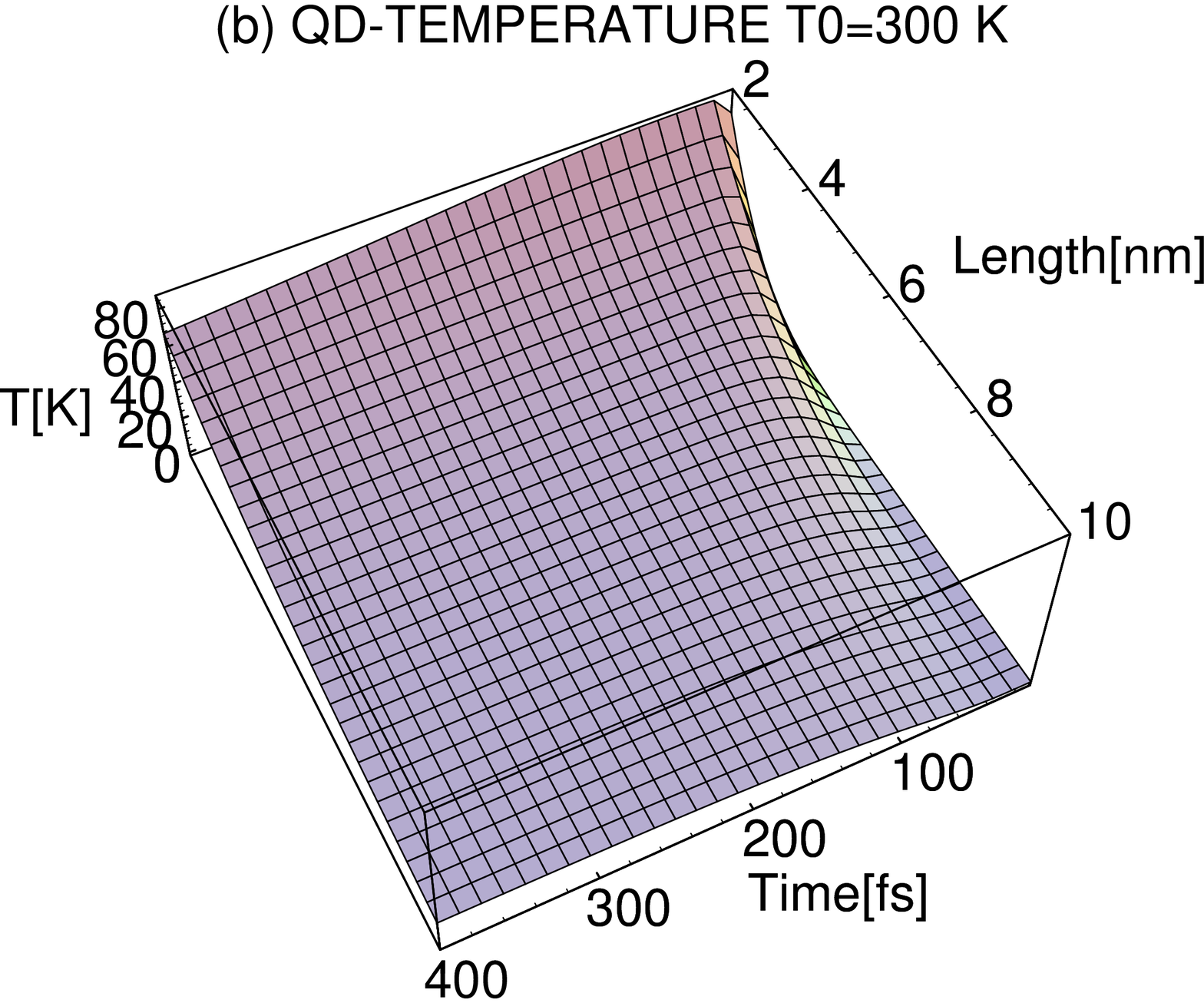}}
\end{picture}
\end{center}
\caption{(a,~b) The same as in Fig.~\ref{fig14.1}(a, b) but for
$\Delta t=\tau$.\label{fig14.2}}
\end{figure}}
In the following we assume that the ambient temperature $T_m$ is
the initial temperature of the nerve cells.

In Figs.~\ref{fig14.1}, \ref{fig14.2} the results of the theoretical calculations for the quantum heat transport on the molecular scale are presented. In Fig.~\ref{fig14.1}(a) the solution of Eq.~(\ref{eq14:1}) in one-dimensional case
\begin{equation}
\tau\frac{\partial^2 T}{\partial t^2}+\frac{\partial T}{\partial t}=D\frac{\partial^2 T}{\partial x^2},\label{eq14.10}
\end{equation}
for the following input parameters $v_h=0.05$ nm/fs, $\tau=44$ fs, $T_0=300$ K (initial temperature) and $\Delta t$ duration of laser pulse $=0.2\tau$ is presented.

In Fig.~\ref{fig14.1}(b) the solution of
quantum
parabolic
heat transport equation (QPT), (Fourier equation)\index{Parabolic equation!for
molecular gas}
\begin{equation}
\frac{\partial T}{\partial t}=D\frac{\partial^2 T}{\partial x^2}\label{eq14:11}
\end{equation}
with the same input parameters is presented.

In Figs.~\ref{fig14.2}(a, b), the solutions of Eqs.~(\ref{eq14:1}), (\ref{eq14:11}) for the
same input parameter but for $\Delta t=\tau$ are presented.

From the analysis of the solutions of hyperbolic and parabolic
quantum heat transport equation, the following conclusions can be
drawn. In the  case of QHT equation  the thermal wave  dominates
the heat  transport for $\Delta t=0.2 \; \tau$. The finite value
of $v_h$ involves the delay time for the response of the molecular
system on the initial temperature change. In the case of QPT,
instantaneous diffusion heat transport is observed. From the
technological point of view, the strong localization of thermal
energy in the front of thermal
wave is very important.%
\section{Electron thermal relaxation in microtubules}
Clusters and aggregates of atoms in the nanometer size (currently
called nanoparticles) are systems intermediate in several
respects, between simple molecules and bulk materials and have
been objects of intensive \linebreak work.

In this paper, we investigate the thermal relaxation phenomena in
the nanoparticles - microtubules in the frame of quantum heat
transport equation. In the book~\cite{5}, the thermal inertia of
materials heated with laser pulses faster than the characteristic
relaxation time was investigated. It was shown, that in the case
of the ultra-short laser pulses the hyperbolic heat conduction
(HHC) must be used. For microtubules the diameters are of the
order of the electron de Broglie wave length. In that case to
description of the transport phenomena quantum heat transport must
be used~\cite{5}.
\begin{equation}
\tau\frac{\partial^2 T}{\partial t^2}+ \frac{\partial T}{\partial
t}=\frac{\hbar}{m}\nabla^2T,\label{eq18:1}
\end{equation}
where $T$ denotes the temperature of the heat carrier, $\tau$ is
the relaxation time and $m$ denotes the mass of heat carrier. The
relaxation time $\tau$ is defined as~\cite{5}
\begin{equation}
\tau=\frac{\hbar}{mv_h^2},\label{eq18:2}
\end{equation}
where $v_h$ is the thermal pulse propagation velocity
\begin{equation}
v_h=\frac{1}{\sqrt{3}}\alpha c.\label{eq18:3}
\end{equation}
In formula~(\ref{eq18:3}) $\alpha$ is the coupling constant (for
electromagnetic interaction $\alpha=e^2/\hbar c$) and $c$ denotes
the light velocity in vacuum. Both parameters $\tau$ and $v_h$
completely characterize the thermal energy transport on the atomic
scale and can be named as ``{\it atomic relaxation time}'' and
{\it ``atomic'' heat velocity}.

Both $\tau$ and $v_h$ are build up from constant of Nature,
$\alpha, c$. Moreover, on the atomic scale there is no shorter
time period than $\tau$ and smaller velocity build from constant
of the Nature. In consequence, one can name $\tau$ and $v_h$ as
{\it elementary relaxation time}  and {\it elementary velocity},
 which characterize heat transport in the elementary
building block of matter, the atom.

In the following, starting with elementary $\tau$ and $v_h$, we
intend to describe thermal relaxation processes in microtubules
which consist of the $N$ parts (molecules) each with elementary
$\tau$ and $v_h$. To that aim, we use the Pauli-Heisenberg
inequality~\cite{5}
\begin{equation}
\Delta r\Delta p\geq N^{\frac{1}{3}}\hbar,\label{eq18:4}
\end{equation}
where $r$ denotes characteristic dimension of the nanoparticle and $p$ is the
momentum of energy carriers. The Pauli-Heisenberg inequality expresses the
basic property of the $N$-fermionic system. In fact, compared to the standard
Heisenberg inequality
\begin{equation}
\Delta r\Delta p\geq\hbar,\label{eq18:5}
\end{equation}
we notice that, in this case the presence of the large number of identical
fermions forces the system either to become spatially more extended for fixed
typical momentum dispension, or to increase its typical momentum dispension for
a~fixed typical spatial extension. We could also say that for a~fermionic
system in its ground state, the average energy per particle increases with the
density of the system.

A picturesque way of interpreting the Pauli-Heisenberg inequality
is to compare Eq.~(\ref{eq18:4}) with Eq.~(\ref{eq18:5}) and to
think of the quantity on the right hand side of it as the
\textit{effective fermionic Planck constant}
\begin{equation}
h^f(N)=N^{\frac13}\hbar.\label{eq18:6}
\end{equation}
We could also say that antisymmetrization, which typifies fermionic amplitudes
amplifies those quantum effects which are affected by Heisenberg inequality.

According to formula~(\ref{eq18:6}), we recalculate the relaxation
time $\tau$, formula~(\ref{eq18:2}) and thermal velocity $v_h$,
formula~(\ref{eq18:3}) for nanoparticle consisting $N$ fermions
\begin{equation}
\hbar\rightarrow\hbar^f (N)=N^{\frac{1}{3}}\hbar\label{eq18:7}
\end{equation}
and obtain
\begin{eqnarray}
v_h^f&=&\frac{e^2}{\hbar^f(N)}=\frac{1}{N^{\frac13}}v_h,\label{eq18:8} \\
\tau^f&=&\frac{\hbar^f}{m(v_h^f)^2}=N\tau.\label{eq18:9}
\end{eqnarray}
Number $N$ parts in nanoparticle (sphere with radius $r$) can be
calculated according to the formula (we assume that density of
nanoparticle does not differ too much from bulk material)
\begin{equation}
N=\frac{\frac{4\pi}{3}r^3\rho AZ}{\mu}\label{eq18:10}
\end{equation}
and for spherical with semiaxes $a, b, c$
\begin{equation}
N=\frac{\frac{4\pi}{3}abc\rho AZ}{\mu},\label{eq18:11}
\end{equation}
where $\rho$ is the density of the nanoparticle, $A$ is the Avogardo number,
$\mu$ is the molecular mass of particles in grams and $Z$ is the number of the
valence electrons.

With formulas~(\ref{eq18:8}) and~(\ref{eq18:9}), we calculated de Broglie wave length
$\lambda_B^f$ and mean free path $\lambda^f_{mfp}$ for nanoparticles
\begin{eqnarray}
\lambda_B^f&=&\frac{\hbar^f}{mv_{th}^f}=N^{\frac{2}{3}}\lambda_B,\label{eq18:12} \\
\lambda_{mfp}^f&=&v_{th}^f\,\tau_{th}^f=N^{\frac{2}{3}}\lambda_{mfp},\label{eq18:13}
\end{eqnarray}
where $\lambda_B$ and $\lambda_{mfp}$ denote the de Broglie wave
length and
mean free path  for heat carriers in nanoparticles.

\section{Quantum transport in microtubules}
Microtubules are essential to cell functions. In neurons
microtubules help and regulate synaptic activity responsible for
learning and cognitive functions. While microtubules have
traditionally been considered to be purely structural elements,
recent evidence has revelated that mechanical, chemical and
electrical signaling and communication function also exist, the
result of microtubule interaction with membrane structures by
linking proteins, ions and voltage fields respectively.

The characteristic dimensions of the microtubules, crystalline
cylinder 10~nm in inner diameter are of the order of the de
Broglie length for electrons in atoms.

When the characteristic length of the structure is of the order of
the de Broglie wave length, then the signaling phenomena must be
described within the quantum transport theory.

For quantum transport phenomena in microtubules we will apply the
equation~(\ref{eq14:1}) with the relaxation time described by
formula~(\ref{eq14:2})
$$
\tau=\frac{2\hbar}{mv^2}=\frac{\hbar}{E}.
$$
The relaxation time is the decoherence time, i.e. time until
collapse of the wave function occurs, when the transition
classical $\to$ quantum phenomena is considered.

In the following we will consider the time $\tau$ for the atomic
and multiatomic phenomena. As was shown in section 2 for atomic
phenomena
\begin{equation}
\tau_a\sim10^{-17}~{\rm s}\label{eq18:14}
\end{equation}
and when we consider multiatomic transport phenomena, with $N$
equal number of agregates involved in transport we have,
formula~(\ref{eq18:9})
\begin{equation}
\tau_N=N\tau_a.\label{eq18:15}
\end{equation}
The Penrose-Hameroff Orchestrated Objective Reduction (Orch OR)
model~\cite{2} proposes that quantum superposition - computation
occurs in nanotubule automata within brain neurons and glia.
Tubulin subumits within microtubules act as qubits, switching
between states on a nanosecond scale, governed by London forces in
hydrophobic pockets. These oscillations are tuned and orchestrated
by microtubule associated proteins (MAPs) providing a feedback
loop between the biological system and the quantum state. These
qubits interact computationally by nonlocal quantum entanglement,
according to the Schr\"odinger equation with preconscious
processing continuing until the threshold for objective reduction
(OR) is reacted $(E=\frac{\hbar}{T})$. At that instant, collapse
occurs, triggering a ``moment of awareness" or a conscious event -
an event that determines particular configurations of Planck scale
experiential geometry and corresponding classical states of
nanotubules automata that regulate synaptic and other neural
functions. A sequence of such events could provide a forward flow
of subjective time and stream of consciousness. Quantum states in
nanotubules may link to those in nanotubules in other neurons and
glia by tunnelling through gap functions, permitting extension of
the quantum state through significant volumes of the brain.

\begin{table}[h]
\caption{}\smallskip
\begin{tabular}[12cm]{|l|c|c|c|c|}\hline
Event&$T$ [ms]&$E$&$N$                 &$\tau$ [ms]\\
     &        &   &number of aggregates&\\ \hline Buddhist
moment&13&$4\cdot10^{15}$ nucleons&$10^{15}$&10\\ of
awareness&&&&\\ \hline Coherent 40Hz
&25&$2\cdot10^{15}$&$10^{15}$&10\\
oscillations&&&&
\\ \hline
EEG alpha rhytm&100&$10^{14}$&$10^{14}$&1 \\
(8 to 12 Hz)&&&&\\ \hline
 Libet's sensory&500&$10^{14}$&$10^{14}$&1 \\
threshold&&&&\\ \hline
\end{tabular}
\end{table}

From $E=\frac{\hbar}{T}$, the size and extension of Orch OR events
that correlate with subjective or neurophysiological description
of conscious events can be calculated. In Table~1 the calculated
$T$ (Penrose-Hameroff) and $\tau-$ formula~(\ref{eq18:15}) are
presented~\cite{6}.

\section{Heisenberg uncertainty principle for thermal phenomena in microtubules} Efficient
conversion of electromagnetic energy to particle energy is of
fundamental importance in many areas of physics. The nature of
intense, short pulse laser interactions with single atoms and
solid targets has been subject of extensive experimental and
theoretical investigation over the last 15~years. Recently, the
interaction of femtosecond laser pulses with $Xe$ clusters was
investigated and strong X-ray emission and multi-keV electron
generation were observed. Such experiments have become possible,
owing to recently developed high peak power lasers which are based
on chirped pulse amplification and are capable of producing
focused light intensity of up to $10^{14} - 10^{19} \, {\rm
Wcm^{-2}}$.

In  intensely irradiated clusters e.g. microtubules, optically and
collisionally ionized electrons undergo rapid collisional heating
for short time ($<$ 1ps) before the cluster disintegrates in the
laser field. Charge separation of the hot electrons inevitably
leads to a~very fast expansion of the cluster ions. Both electrons
and ions ultimately reach a~velocity given by the speed of sound
of the cluster plasma.

When the intense laser pulse interacts with atomic clusters
ionization to very high charge states is observed. The high
Coulomb field certainly influences the thermal processes in
clusters. In the chapter, the new QHT~equation is formulated in
which the external --- not only thermal forces are included. The
solution of the new QHT for Cauchy boundary conditions will be
derived. The condition for the distortionless propagations of the
thermal wave will be formulated.

Now, we develop the generalized quantum heat transport equation
which includes the potential term. In this way, we use the analogy
between the Schr\"odinger equation and quantum heat transport
equations. Let us consider, for the moment, the parabolic heat
transport equation with the second derivative term omitted
\begin{equation}
\frac{\partial T}{\partial t}=\frac{\hbar}{m}\nabla^2T.\label{eq10:2}
\end{equation}
When the real time $t\rightarrow\frac{it}{2}$  and $T\rightarrow\Psi$,
Eq.~(\ref{eq10:2}) has the form of a~free Schr\"odinger equation
\begin{equation}
i\hbar\frac{\partial \Psi}{\partial
t}=-\frac{\hbar^2}{2m}\nabla^2\Psi.\label{eq10:3}
\end{equation}
The complete Schr\"odinger equation\index{Schr\"odinger equation} has the form
\begin{equation}
i\hbar\frac{\partial \Psi}{\partial t}=
-\frac{\hbar^2}{2m}\nabla^2\Psi+V\Psi,\label{eq10:4}
\end{equation}
where $V$ denotes the potential energy.
When we go back to real time $t\rightarrow-2it$ and $\Psi\rightarrow T$, the new
parabolic quantum heat transport is obtained
\begin{equation}
\frac{\partial T}{\partial
t}=\frac{\hbar}{m}\nabla^2T-\frac{2V}{\hbar}T.\label{eq10:5}
\end{equation}
Equation~(\ref{eq10:5}) describes the quantum heat transport for $\Delta t>\tau$.
For heat transport initiated by ultra-short laser pulses, when $\Delta t \leq \tau$
one obtains the generalized quantum hyperbolic heat transport
equation
\begin{equation}
\tau\frac{\partial^2 T}{\partial t^2}+\frac{\partial T}{\partial
t}=\frac{\hbar}{m}\nabla^2T-\frac{2V}{\hbar}T.\label{eq10:6}
\end{equation}
Considering that $\tau=\hbar/{mv^2}$, Eq.~(\ref{eq10:6}) can be
written as follows:
\begin{equation}
\frac{1}{v^2}\frac{\partial^2 T}{\partial t^2}+\frac{m}{\hbar}\frac{\partial
T}{\partial t}+\frac{2Vm}{\hbar}T=\nabla^2T.\label{eq10:7}
\end{equation}
Equation~(\ref{eq10:7}) describes the heat flow when apart from the temperature
gradient, the potential energy $V$ operates.

In the following, we consider the one-dimensional heat transfer phenomena,~i.e.
\begin{equation}
\frac{1}{v^2}\frac{\partial^2 T}{\partial t^2}+\frac{m}{\hbar}\frac{\partial
T}{\partial t}+\frac{2Vm}{\hbar}T=\frac{\partial^2 T}{\partial x^2}.\label{eq10:8}
\end{equation}
For quantum heat transfer equation~(\ref{eq10:8}), we seek solution in the form
\begin{equation}
T(x, t)=e^{-t/{2\tau}}u(x, t).\label{eq10:9}
\end{equation}
After substitution of Eq.~(\ref{eq10:9}) into Eq.~(\ref{eq10:8}), one obtains
\begin{equation}
\frac{1}{v^2}\frac{\partial^2 u}{\partial t^2}-\frac{\partial^2 u}
{\partial x^2}+q\,u(x,t)=0,\label{eq10:10}
\end{equation}
where
\begin{equation}
q=\frac{2Vm}{\hbar^2}-\left(\frac{mv}{2\hbar}\right)^2.\label{eq10:11}
\end{equation}
In the following, we will consider the constant potential energy $V=V_0$. The~general
solution of Eq.~(\ref{eq10:10}) for Cauchy boundary conditions,
\begin{equation}
u(x, 0)=f(x), \qquad
\left.\frac{\partial u(x,t)}{\partial t}\right\vert_{t=0}=F(x),\label{eq10:12}
\end{equation}
has the form~\cite{5}
\begin{equation}
u(x, t)=\frac{f(x-vt)+f(x+vt)}{2}+\frac{1}{2}\int^{x+vt}_{x-vt}
\Phi(x, t, z)dz,\label{eq10:13}
\end{equation}
where
\begin{eqnarray}
\Phi(x, t, z)&=&\frac1vF(z)J_0\left(\frac{b}{v}\sqrt{(z-x)^2-v^2t^2}\right)+
btf(z)\frac{J_0^{\prime}\left(\displaystyle\frac{b}{v}\sqrt{\left(z-x\right)^2-v^2t^2}\right)}
{\sqrt{\left(z-x\right)^2-v^2t^2}},\nonumber\\
b&=&\left(\frac{mv^2}{2\hbar}\right)^2-\frac{2Vm}{\hbar^2}v^2\label{eq10:14}
\end{eqnarray}
and $J_0(z)$ denotes the Bessel function of the first kind.
Considering formulas~(\ref{eq10:9}), (\ref{eq10:10}), (\ref{eq10:11}) the solution of
Eq.~(\ref{eq10:8}) describes the propagation of the distorted thermal quantum
waves with characteristic lines $x=\pm vt$. We can define the distortionless
thermal wave as the wave which preserves the shape in the field of the
potential energy $V_0$. The condition for conserving the shape can be
formulated as
\begin{equation}
q=\frac{2Vm}{\hbar^2}-\left(\frac{mv}{2\hbar}\right)^2.\label{eq10:15}
\end{equation}
When Eq.~(\ref{eq10:15}) holds,  Eq.~(\ref{eq10:10}) has the form
\begin{equation}
\frac{\partial^2 u(x, t)}{\partial t^2}=v^2\frac{\partial^2  u}{\partial x^2}.
\label{eq10:16}
\end{equation}
Equation~(\ref{eq10:16}) is the quantum wave equation with the solution (for
Cauchy boundary conditions~(\ref{eq10:12}))
\begin{equation}
u(x, t)=\frac{f(x-vt)+f(x+vt)}{2}+\frac{1}{2v}\int^{x+vt}_{x-vt}F(z)dz.
\label{eq10:17}
\end{equation}
It is quite interesting to observe, that condition~(\ref{eq10:15}) has an analog
 in~the  classical theory of the electrical transmission line. In the
context  of~the
transmission of an electromagnetic field, the condition $q=0$ describes the
Heaviside distortionless line. Eq.~(\ref{eq10:15})~---~the
distortionless condition~---~can be written as
\begin{equation}
V_0\tau\sim\hbar,\label{eq10:18}
\end{equation}
We can conclude, that in the presence of the potential energy
$V_0$ one can observe the undisturbed quantum thermal wave only
when \textit{the Heisenberg uncertainty relation for thermal
processes}~(\ref{eq10:18}) is fulfilled.

The generalized  quantum heat transport equation
(GQHT)~(\ref{eq10:8})  leads to the generalized Schr\"odinger
equation. After the substitution $t\rightarrow it/2$,
$T\rightarrow \Psi$ in~Eq.~(\ref{eq10:8}), one obtains the
generalized Schr\"odinger equation~(GSE)
\begin{equation}
i\hbar \frac{\partial \Psi}{\partial t}=-\frac{\hbar^2}{2m}\nabla^2
\Psi+V\Psi-2\tau \hbar \frac{\partial^2 \Psi}{\partial t^2}.\label{eq10:19}
\end{equation}
Considering that $\tau=\hbar/mv^2=\hbar/m\alpha^2c^2$ ($\alpha=1/137$ is the
fine-structure constant for electromagnetic interactions) Eq.~(\ref{eq10:19})
can be written as
\begin{equation}
i\hbar\frac{\partial \Psi}{\partial t}=-\frac{\hbar^2}{2m}\nabla^2 \Psi+V\Psi-
\frac{2\hbar^2}{m\alpha^2c^2}\frac{\partial^2 \Psi}{\partial t^2}.\label{eq10:20}
\end{equation}
One can conclude, that for time period $\Delta
t<\hbar/m\alpha^2c^2\sim10^{-17}\;
{\rm s}$ the description of quantum phenomena needs some revision. On the other
hand, for~$\Delta t>10^{-17}$ in GSE the second derivative term can be omitted
and as the result the SE is obtained,~i.e.
\begin{equation}
i\hbar\frac{\partial \Psi}{\partial t}=-\frac{\hbar^2}{2m}\nabla^2\Psi+V\Psi.
\label{eq10:21}
\end{equation}
It is quite interesting to observe, that GSE  was discussed also
in the context of the sub-quantal phenomena.

Concluding, a~study of the interactions of the attosecond laser
pulses with matter can shed the light on the applicability of the
SE to the study of the ultra-short sub-quantal phenomena.

 The
structure of the Eq.~(\ref{eq10:10}) depends on the sign of the
parameter $q$. For quantum heat transport phenomena with electrons
as the heat carriers parameter $q$ is the function of potential
barrier height $(V_0)$ and velocity $v$. Considering that velocity
$v$ equals
\begin{equation}
v=\frac1{\sqrt{3}}\alpha c=1.26 \, {\rm \frac{nm}{fs}},\label{eq11:7}
\end{equation}
parameter $q$ can be calculated for typical barrier height
$V_0\geq0$. In Fig.~\ref{fig11.1} the parameter $q$ as the
function of $V_0$ is calculated. For $q<0$, i.e., when $V_0<$1.25
eV, Eq.~(\ref{eq10:10}) is the {\it modified Klein-Gordon
equation}.

{\linespread{1}
\begin{figure}[h]
\begin{center}
\begin{picture}(341,211)
\put(20,-100) {\includegraphics{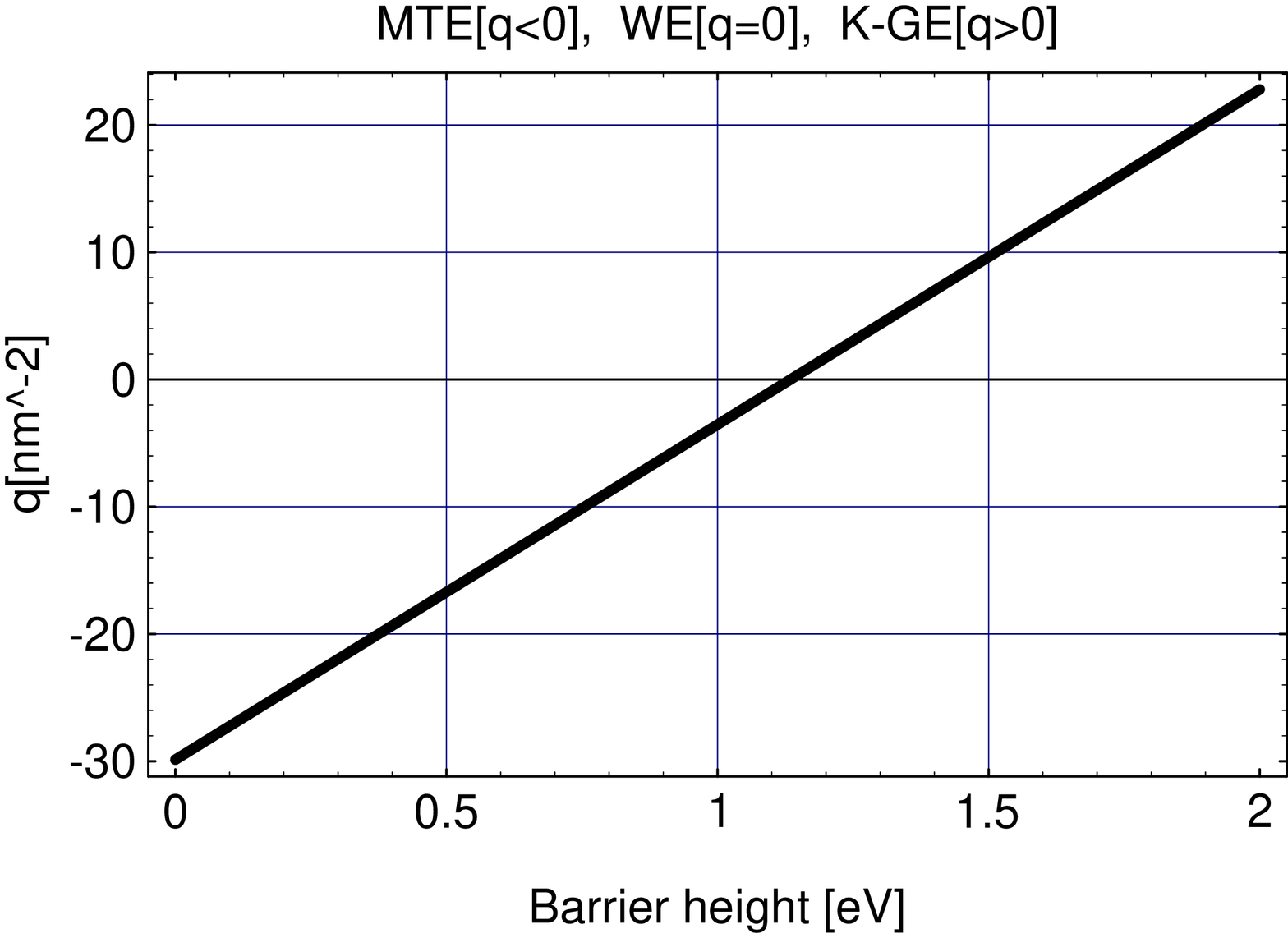}}
\vspace{8cm}
\end{picture}
\end{center}
\caption{Parameter $q$ (formula~(\ref{eq10:11})) as the function of the barrier
height~(eV).\label{fig11.1}}
\end{figure}}
For Cauchy initial condition
\begin{equation}
u(x,o)=f(x), \qquad \frac{\partial u(x, o)}{\partial t}=g(x),\label{eq11:8}
\end{equation}
the solution of the Eq.~(\ref{eq10:10}) has the form
\begin{eqnarray}
u(x, t)&=&\frac{f(x-vt)+f(x+vt)}2 \label{eq11:9} \\*
&&\mbox{}+\frac1{2v}\int_{x-vt}^{x+vt}
g(\zeta)I_0\left[\sqrt{-q(v^2t^2-(x-\zeta)^2)}\right]d\zeta \nonumber \\*
&&\mbox{}+\frac{(v\sqrt{-q})t}2\int_{x-vt}^{x+vt}f(\zeta)
\frac{I_1\left[\sqrt{-q(v^2t^2-(x-\zeta)^2)}\right]}
{\sqrt{v^2t^2-(x-\zeta)^2}}d\zeta.\nonumber
\end{eqnarray}
When $q>0$  Eq.~(\ref{eq10:10}) is the {\it Klein-Gordon
equation}~(K-G) well known from application to elementary particle
and nuclear physics.

For Cauchy initial condition~(\ref{eq11:8}), the solution of
(K-G) equation can be written as
\begin{eqnarray}
u(x, t)&=&\frac{f(x-vt)+f(x+vt)}2 \label{eq11:10} \\*
&&\mbox{}+\frac1{2v}\int_{x-vt}^{x+vt}g(\zeta)J_0
\left[\sqrt{q(v^2t^2-(x-\zeta)^2)}\right]d\zeta \nonumber\\*
&&\mbox{}-\frac{v\sqrt{q}t}2\int_{x-vt}^{x+vt}f(\zeta)
\frac{J_0^\prime\left[\sqrt{q(v^2t^2-(x-\zeta)^2)}\right]}{\sqrt{v^2t^2-(x-\zeta)^2}}
d\zeta.\nonumber
\end{eqnarray}

Both solutions~(\ref{eq11:9}) and (\ref{eq11:10}) exhibit the
domains of dependence and influence on {\it modified Klein-Gordon
equation} and {\it Klein--Gordon equation}. These domains, which
characterize the maximum speed at which thermal disturbance travel
are determined by the principal parts of the given equation (i.e.,
the second derivative terms) and do not depend on the lower order
terms. It can be concluded that these equation and the wave
equation (for $m=0$) have identical domains of dependence and
influence.

\section{Thermal wave packets induced by attosecond laser pulses}
Equation{\nobreakspace}(\ref{eq10:10}) is the modified
Klein-Gordon equation which can be written as
\begin{equation}
\bar{\Box}u-\left(\frac{mv}{2\hbar}\right)^2u(x,t)=0,\label{eq5:5}
\end{equation}
where d'Alembert operator $\bar{\Box}$ is equal
\begin{equation}
\bar{\Box}=\frac{1}{v^2}\frac{\partial}{\partial
t^2}-\frac{\partial}{\partial x^2}.\label{eq5:6}
\end{equation}
The ordinary Klein-Gordon equation for the particle with mass $m$
and velocity $v$ is of the form{\nobreakspace}\cite{5}
\begin{equation}
\bar{\Box}u+\left(\frac{m_0}{2\hbar}\right)^2u=0.\label{eq5:7}
\end{equation}
Equation{\nobreakspace}(\ref{eq5:5}) can be split into its real
and imaginary parts. Putting for $u(x,t)$
$$
 u(x,t)=\Re (t,x)\exp \left[ \frac{i}{\hbar } S(t,x)\right]
$$
one obtains
\begin{equation}
\eta ^{ab} (\partial _{a} S)(\partial _{b} S)=\hbar ^{2}
\frac{\bar{\Box}\Re
}{\Re}-\left(\frac{mv}{2}\right)^2,\label{eq5:8}
\end{equation}
where
$$\eta _{ab} ={\rm diag}(1,-1,-1,-1),\quad \quad a,b=1,2,3,4.$$

We use Mackinnon's suggestion{\nobreakspace}~\cite{6} therefore we
look for solutions that satisfy the equation
\begin{equation}
\frac{\bar{\Box}\Re
}{\Re}=\left(\frac{mv}{\sqrt{2}\hbar}\right)^2.\label{eq5:9}
\end{equation}
In this way Eq.{\nobreakspace}(\ref{eq5:8}) becomes
\begin{equation}
\eta ^{ab} (\partial _{a} S)(\partial _{b} S)=m^{2} v^{2} =P_{\mu
} P^{\mu }.\label{eq5:10}
\end{equation}
If the velocity $v$ is constant, we have
from{\nobreakspace}(\ref{eq5:8})
\begin{equation}
S=-P^{\mu } P_{\mu }\label{eq5:11}
\end{equation}
and the de Broglie relation $P^{v} =\hbar K^{v} $ hold where
$K^{v} $
 is the wave number and $P^{v} $  is the classical relativistic four momentum,
$P^{v} =\left( \frac{E}{v} \vec{p} \right) $. In
paper{\nobreakspace}[6] L.{\nobreakspace}Mackinnon constructs a
wave packet considering that a wave $\Phi =\Phi _{0} \exp \left[
i\omega t\right] $
 of frequency
$\omega =\frac{mc^{2} }{\hbar } $
 is associated with a particle of rest mass $m$ and that for
an observer moving with a constant velocity $v$
 with respect to the particle, the associated wave (be means
of a Lorentz transformation) acquires the form $\Phi =\Phi _{0}
\exp \left[ i\omega \gamma \left( t-\frac{v}{c^{2} } x\right)
\right] $. Mackinnon showed the wave packet is compatible with the
basic experiments of quantum mechanics and does not spread in time
the found
\begin{equation}
\Re =A\frac{\sin[gr]}{gr},\label{eq5:12}
\end{equation}
where $A=$ constant and $g=\frac{mv}{\sqrt{2} \hbar } $, and
$g=\frac{m\alpha c}{\sqrt{2} \hbar}  $
 and
\begin{equation}
r=\gamma (x-vt)\label{eq5:13}
\end{equation}
is the distance from the particle portion, so that
\begin{equation}
u(x,t)=A\frac{\sin gr}{gr} \exp \left[ -\frac{i}{\hbar }
(Et-px)\right] \label{eq5:14}
\end{equation}
is the Lorentz boost of the solution
$$\Phi =A\frac{\sin gr'}{gr'} \exp \left[ -i\omega t'\right],$$
where $r'=x$. This solution was first found by de
Broglie{\nobreakspace}~\cite{8} and also used in the stochastic
interpretation of quantum mechanics by Vigier and
Gueret{\nobreakspace}\cite{9}.


\newpage
\begin{thebibliography}{99}
\bibitem{1}J.~G.~Lambert et al., \textit{Cell. Mol. Biol.}, \textbf{45,} (1999),
p.~905.
\bibitem{2}R.~D.~Vale et al., \textit{Cell},  (1985), p.~449.
\bibitem{3}D.~G.~Weiss et al., \textit{Cell Motil. Cytoskeleton},
\textbf{6}, (1986), p.~128.
\bibitem{4}D.~A.~Smith, R.~M.~Simmons, \textit{Biophysical
Journal}, \textbf{80}, (2001), p.~46.
\bibitem{5}M. Kozlowski, J. Marciak-Kozlowska, \textit{From Quarks to Bulk
Matter}, Hadronic Press, 2001.
\bibitem{6}S.~R.~Hameroff, J.~Tuszy\'nski, {Conscious events
as orchestrated spacetime selections}, \textit{Journal of
Consciousness Studies}, \textbf{3}, (1996), p.~36.
\bibitem{7}L.~Mackinnon, \textit{Letter at Nuovo Cimento}, \textbf{32}, (1981),
p.~311.
\bibitem{8}L. de Broglie, \textit{C. R. Acad. Sci}., \textbf{180}, (1925), p. 498.
\bibitem{9}J.~P.~Vigier and Ph.~Gueret, \textit{Found. Phys}., \textbf{12},
(1982), p.~1057.
\bibitem{10}G.~Cappelo et al., \textit{Phys. Rev.}, \textbf{E68},
(2003), 021907-1.
\end{thebibliography}
\end{document}